
\def\nuc #1,#2,#3{{\sl Nucl. Phys.} {\bf B#1} (#2) #3}
\def\pr #1,#2,#3{{\sl Phys. Rev.} {\bf #1} (#2) #3}
\def\part{\partial}
\def\al{\alpha}

\def\be{\beta}
\def\gam{\gamma}
\def\ga{\gamma}
\def\la{\lambda}
\def\Lam{\Lambda}
\def\Gam{\Gamma}
\def\ie{{\it i.e.,}\ }

\def\ee{\hbox{e}}
\def\dd{\hbox{d}}
\def\dD{\hbox{D}}
\def\ttt{{2\over3}}
\def\ee{\hbox{e}}
\def\dD{\hbox{D}}
\def\vw{\hbox{vol.(Weyl)}}
\def\eg{{\it e.g.},}
\def\pref#1{${}^{\cite{#1}}$}
\def\hg{{\hat g}}
\def\Del{\Delta}
\def\Gam{\Gamma}
\def\ssc{\scriptscriptstyle}
\def\sc{\scriptstyle}
\catcode`@=11
\expandafter\ifx\csname inp@t\endcsname\relax\let\inp@t=\input
\def\input#1 {\expandafter\ifx\csname #1IsLoaded\endcsname\relax
\inp@t#1%
\expandafter\def\csname #1IsLoaded\endcsname{(#1 was previously loaded)}
\else\message{\csname #1IsLoaded\endcsname}\fi}\fi
\catcode`@=12

\font\twelverm=cmr12			\font\twelvei=cmmi12
\font\twelvesy=cmsy10 scaled 1200	\font\twelveex=cmex10 scaled 1200
\font\twelvebf=cmbx12			\font\twelvesl=cmsl12
\font\twelvett=cmtt12			\font\twelveit=cmti12
\font\twelvesc=cmcsc10 scaled 1200	\font\twelvesf=cmss12
\skewchar\twelvei='177			\skewchar\twelvesy='60


\def\twelvepoint{\normalbaselineskip=12.4pt plus 0.1pt minus 0.1pt
  \abovedisplayskip 12.4pt plus 3pt minus 9pt
  \belowdisplayskip 12.4pt plus 3pt minus 9pt
  \abovedisplayshortskip 0pt plus 3pt
  \belowdisplayshortskip 7.2pt plus 3pt minus 4pt
  \smallskipamount=3.6pt plus1.2pt minus1.2pt
  \medskipamount=7.2pt plus2.4pt minus2.4pt
  \bigskipamount=14.4pt plus4.8pt minus4.8pt
  \def\rm{\fam0\twelverm}          \def\it{\fam\itfam\twelveit}%
  \def\sl{\fam\slfam\twelvesl}     \def\bf{\fam\bffam\twelvebf}%
  \def\mit{\fam 1}                 \def\cal{\fam 2}%
  \def\sc{\twelvesc}		   \def\tt{\twelvett}
  \def\sf{\twelvesf}
  \textfont0=\twelverm   \scriptfont0=\tenrm   \scriptscriptfont0=\sevenrm
  \textfont1=\twelvei    \scriptfont1=\teni    \scriptscriptfont1=\seveni
  \textfont2=\twelvesy   \scriptfont2=\tensy   \scriptscriptfont2=\sevensy
  \textfont3=\twelveex   \scriptfont3=\twelveex  \scriptscriptfont3=\twelveex
  \textfont\itfam=\twelveit
  \textfont\slfam=\twelvesl
  \textfont\bffam=\twelvebf \scriptfont\bffam=\tenbf
  \scriptscriptfont\bffam=\sevenbf
  \normalbaselines\rm}



\def\beginlinemode{\endmode
  \begingroup\parskip=0pt \obeylines\def\\{\par}\def\endmode{\par\endgroup}}
\def\beginparmode{\endmode
  \begingroup \def\endmode{\par\endgroup}}
\let\endmode=\par
{\obeylines\gdef\
{}}
\def\singlespace{\baselineskip=\normalbaselineskip}

\def\oneandahalfspace{\baselineskip=\normalbaselineskip
  \multiply\baselineskip by 3 \divide\baselineskip by 2}
\def\doublespace{\baselineskip=\normalbaselineskip \multiply\baselineskip by 2}

\newcount\firstpageno
\firstpageno=2
\footline={\ifnum\pageno<\firstpageno{\hfil}\else{\hfil\twelverm\folio\hfil}\fi}
\def\toppageno{\global\footline={\hfil}\global\headline
  ={\ifnum\pageno<\firstpageno{\hfil}\else{\hfil\twelverm\folio\hfil}\fi}}
\let\rawfootnote=\footnote		
\def\footnote#1#2{{\rm\singlespace\parindent=0pt\parskip=0pt
  \rawfootnote{#1}{#2\hfill\vrule height 0pt depth 6pt width 0pt}}}
\def\raggedcenter{\leftskip=4em plus 12em \rightskip=\leftskip
  \parindent=0pt \parfillskip=0pt \spaceskip=.3333em \xspaceskip=.5em
  \pretolerance=9999 \tolerance=9999
  \hyphenpenalty=9999 \exhyphenpenalty=9999 }
\def\dateline{\rightline{\ifcase\month\or
  January\or February\or March\or April\or May\or June\or
  July\or August\or September\or October\or November\or December\fi
  \space\number\year}}
\def\received{\vskip 3pt plus 0.2fill
 \centerline{\sl (Received\space\ifcase\month\or
  January\or February\or March\or April\or May\or June\or
  July\or August\or September\or October\or November\or December\fi
  \qquad, \number\year)}}


\hsize=6.5truein
\hoffset=0pt
\vsize=8.9truein
\voffset=0pt
\parskip=\medskipamount
\def\\{\cr}
\twelvepoint		
\doublespace		
\overfullrule=0pt	




\def\title			
  {\null\vskip 3pt plus 0.2fill
   \beginlinemode \doublespace \raggedcenter \bf}

\def\author			
  {\vskip 3pt plus 0.2fill \beginlinemode
   \singlespace \raggedcenter\sc}

\def\affil			
  {\vskip 3pt plus 0.1fill \beginlinemode
   \oneandahalfspace \raggedcenter \sl}

\def\abstract			
  {\vskip 3pt plus 0.3fill \beginparmode
   \baselineskip=16truept ABSTRACT: }

\def\endtitlepage		
  {\endpage			
   \body}
\let\endtopmatter=\endtitlepage

\def\body			
  {\beginparmode}		

\def\head#1{			
  \goodbreak\vskip 0.5truein	
  {\immediate\write16{#1}
   \raggedcenter \uppercase{#1}\par}
   \nobreak\vskip 0.25truein\nobreak}

\def\beginitems{
\par\medskip\bgroup\def\i##1 {\item{##1}}\def\ii##1 {\itemitem{##1}}
\leftskip=36pt\parskip=0pt}
\def\enditems{\par\egroup}

\def\beneathrel#1\under#2{\mathrel{\mathop{#2}\limits_{#1}}}

\def\refto#1{$^{#1}$}		

\def\references			
  {\head{References}		
   \beginparmode
   \frenchspacing \parindent=0pt \leftskip=1truecm
   \parskip=3pt plus 3pt \baselineskip=17truept %
   \everypar{\hangindent=\parindent}}

\gdef\refis#1{\item{#1.\ }}			

\gdef\journal#1, #2, #3, 1#4#5#6{		
    {\sl #1~}{\bf #2}, #3 (1#4#5#6)}		

\def\nuc #1,#2,#3{{\sl Nucl. Phys.} {\bf B#1} (#2) #3}
\def\pr #1,#2,#3{{\sl Phys. Rev.} {\bf #1} (#2) #3}

\def\endreferences{\body}

\def\figurecaptions		
  {\endpage
   \beginparmode
   \head{Figure Captions}
}

\def\endpage			
  {\vfill\eject}

\def\endpaper			
  {\endmode\vfill\supereject}


\def\heading				
  {\vskip 0.5truein plus 0.1truein	
   \beginparmode \def\\{\par} \parskip=0pt \singlespace \raggedcenter}

\def\subheading				
  {\vskip 0.25truein plus 0.1truein	
   \beginlinemode \singlespace \parskip=0pt \def\\{\par}\raggedcenter}

\def\tag#1$${\eqno(#1)$$}

\def\align#1$${\eqalign{#1}$$}

\def\aligntag#1$${\gdef\tag##1\\{&(##1)\cr}\eqalignno{#1\\}$$
  \gdef\tag##1$${\eqno(##1)$$}}

\def\endaligntag{}

\def\overset #1\to#2{{\mathop{#2}\limits^{#1}}}
\def\underset#1\to#2{{\let\next=#1\mathpalette\undersetpalette#2}}
\def\undersetpalette#1#2{\vtop{\baselineskip0pt
\ialign{$\mathsurround=0pt #1\hfil##\hfil$\crcr#2\crcr\next\crcr}}}


\def\ref#1{Ref.~#1}			
\def\Ref#1{Ref.~#1}			
\def\[#1]{[\cite{#1}]}
\def\cite#1{{#1}}
\def\(#1){(\call{#1})}
\def\call#1{{#1}}
\def\taghead#1{}
\def\frac#1#2{{#1 \over #2}}

\def\12{{1\over2}}
\def\eg{{\it e.g.,\ }}

\def\ie{{\it i.e.,\ }}

\def\sla{\raise.15ex\hbox{$/$}\kern-.57em}
\def\leaderfill{\leaders\hbox to 1em{\hss.\hss}\hfill}
\def\twiddle{\lower.9ex\rlap{$\kern-.1em\scriptstyle\sim$}}
\def\bigtwiddle{\lower1.ex\rlap{$\sim$}}
\def\gtwid{\mathrel{\raise.3ex\hbox{$>$\kern-.75em\lower1ex\hbox{$\sim$}}}}
\def\ltwid{\mathrel{\raise.3ex\hbox{$<$\kern-.75em\lower1ex\hbox{$\sim$}}}}
\def\square{\kern1pt\vbox{\hrule height 1.2pt\hbox{\vrule width 1.2pt\hskip 3pt
   \vbox{\vskip 6pt}\hskip 3pt\vrule width 0.6pt}\hrule height 0.6pt}\kern1pt}
\def\tdot#1{\mathord{\mathop{#1}\limits^{\kern2pt\ldots}}}

\def\pmb#1{\setbox0=\hbox{#1}%
  \kern-.025em\copy0\kern-\wd0
  \kern  .05em\copy0\kern-\wd0
  \kern-.025em\raise.0433em\box0 }

\catcode`@=11
\newcount\tagnumber\tagnumber=0

\immediate\newwrite\eqnfile
\newif\if@qnfile\@qnfilefalse
\def\write@qn#1{}
\def\writenew@qn#1{}
\def\w@rnwrite#1{\write@qn{#1}\message{#1}}
\def\@rrwrite#1{\write@qn{#1}\errmessage{#1}}

\def\taghead#1{\gdef\t@ghead{#1}\global\tagnumber=0}
\def\t@ghead{}

\expandafter\def\csname @qnnum-3\endcsname
  {{\t@ghead\advance\tagnumber by -3\relax\number\tagnumber}}
\expandafter\def\csname @qnnum-2\endcsname
  {{\t@ghead\advance\tagnumber by -2\relax\number\tagnumber}}
\expandafter\def\csname @qnnum-1\endcsname
  {{\t@ghead\advance\tagnumber by -1\relax\number\tagnumber}}
\expandafter\def\csname @qnnum0\endcsname
  {\t@ghead\number\tagnumber}
\expandafter\def\csname @qnnum+1\endcsname
  {{\t@ghead\advance\tagnumber by 1\relax\number\tagnumber}}
\expandafter\def\csname @qnnum+2\endcsname
  {{\t@ghead\advance\tagnumber by 2\relax\number\tagnumber}}
\expandafter\def\csname @qnnum+3\endcsname
  {{\t@ghead\advance\tagnumber by 3\relax\number\tagnumber}}

\def\equationfile{%
  \@qnfiletrue\immediate\openout\eqnfile=\jobname.eqn%
  \def\write@qn##1{\if@qnfile\immediate\write\eqnfile{##1}\fi}
  \def\writenew@qn##1{\if@qnfile\immediate\write\eqnfile
    {\noexpand\tag{##1} = (\t@ghead\number\tagnumber)}\fi}
}

\def\callall#1{\xdef#1##1{#1{\noexpand\call{##1}}}}
\def\call#1{\each@rg\callr@nge{#1}}

\def\each@rg#1#2{{\let\thecsname=#1\expandafter\first@rg#2,\end,}}
\def\first@rg#1,{\thecsname{#1}\apply@rg}
\def\apply@rg#1,{\ifx\end#1\let\next=\relax%
\else,\thecsname{#1}\let\next=\apply@rg\fi\next}

\def\callr@nge#1{\calldor@nge#1-\end-}
\def\callr@ngeat#1\end-{#1}
\def\calldor@nge#1-#2-{\ifx\end#2\@qneatspace#1 %
  \else\calll@@p{#1}{#2}\callr@ngeat\fi}
\def\calll@@p#1#2{\ifnum#1>#2{\@rrwrite{Equation range #1-#2\space is bad.}
\errhelp{If you call a series of equations by the notation M-N, then M and
N must be integers, and N must be greater than or equal to M.}}\else%
 {\count0=#1\count1=#2\advance\count1
by1\relax\expandafter\@qncall\the\count0,%
  \loop\advance\count0 by1\relax%
    \ifnum\count0<\count1,\expandafter\@qncall\the\count0,%
  \repeat}\fi}

\def\@qneatspace#1#2 {\@qncall#1#2,}
\def\@qncall#1,{\ifunc@lled{#1}{\def\next{#1}\ifx\next\empty\else
  \w@rnwrite{Equation number \noexpand\(>>#1<<) has not been defined yet.}
  >>#1<<\fi}\else\csname @qnnum#1\endcsname\fi}

\let\eqnono=\eqno
\def\eqno(#1){\tag#1}
\def\tag#1$${\eqnono(\displayt@g#1 )$$}

\def\aligntag#1\endaligntag
  $${\gdef\tag##1\\{&(##1 )\cr}\eqalignno{#1\\}$$
  \gdef\tag##1$${\eqnono(\displayt@g##1 )$$}}

\def\eqalignno#1{\displ@y \tabskip\centering
  \halign to\displaywidth{\hfil$\displaystyle{##}$\tabskip\z@skip
    &$\displaystyle{{}##}$\hfil\tabskip\centering
    &\llap{$\displayt@gpar##$}\tabskip\z@skip\crcr
    #1\crcr}}

\def\displayt@gpar(#1){(\displayt@g#1 )}

\def\displayt@g#1 {\rm\ifunc@lled{#1}\global\advance\tagnumber by1
        {\def\next{#1}\ifx\next\empty\else\expandafter
        \xdef\csname @qnnum#1\endcsname{\t@ghead\number\tagnumber}\fi}%
  \writenew@qn{#1}\t@ghead\number\tagnumber\else
        {\edef\next{\t@ghead\number\tagnumber}%
        \expandafter\ifx\csname @qnnum#1\endcsname\next\else
        \w@rnwrite{Equation \noexpand\tag{#1} is a duplicate number.}\fi}%
  \csname @qnnum#1\endcsname\fi}

\def\ifunc@lled#1{\expandafter\ifx\csname @qnnum#1\endcsname\relax}

\let\@qnend=\end\gdef\end{\if@qnfile
\immediate\write16{Equation numbers written on []\jobname.EQN.}\fi\@qnend}

\catcode`@=12

\catcode`@=11
\newcount\r@fcount \r@fcount=0
\newcount\r@fcurr
\immediate\newwrite\reffile
\newif\ifr@ffile\r@ffilefalse
\def\w@rnwrite#1{\ifr@ffile\immediate\write\reffile{#1}\fi\message{#1}}

\def\writer@f#1>>{}
\def\referencefile{
  \r@ffiletrue\immediate\openout\reffile=\jobname.ref%
  \def\writer@f##1>>{\ifr@ffile\immediate\write\reffile%
    {\noexpand\refis{##1} = \csname r@fnum##1\endcsname = %
     \expandafter\expandafter\expandafter\strip@t\expandafter%
     \meaning\csname r@ftext\csname r@fnum##1\endcsname\endcsname}\fi}%
  \def\strip@t##1>>{}}

\def\citeall#1{\xdef#1##1{#1{\noexpand\cite{##1}}}}
\def\cite#1{\each@rg\citer@nge{#1}}	

\def\each@rg#1#2{{\let\thecsname=#1\expandafter\first@rg#2,\end,}}
\def\first@rg#1,{\thecsname{#1}\apply@rg}	
\def\apply@rg#1,{\ifx\end#1\let\next=\relax
\else,\thecsname{#1}\let\next=\apply@rg\fi\next}

\def\citer@nge#1{\citedor@nge#1-\end-}	
\def\citer@ngeat#1\end-{#1}
\def\citedor@nge#1-#2-{\ifx\end#2\r@featspace#1 
  \else\citel@@p{#1}{#2}\citer@ngeat\fi}	
\def\citel@@p#1#2{\ifnum#1>#2{\errmessage{Reference range #1-#2\space is bad.}%
    \errhelp{If you cite a series of references by the notation M-N, then M and
    N must be integers, and N must be greater than or equal to M.}}\else%
 {\count0=#1\count1=#2\advance\count1
by1\relax\expandafter\r@fcite\the\count0,%
  \loop\advance\count0 by1\relax
    \ifnum\count0<\count1,\expandafter\r@fcite\the\count0,%
  \repeat}\fi}

\def\r@featspace#1#2 {\r@fcite#1#2,}	
\def\r@fcite#1,{\ifuncit@d{#1}
    \newr@f{#1}%
    \expandafter\gdef\csname r@ftext\number\r@fcount\endcsname%
                     {\message{Reference #1 to be supplied.}%
                      \writer@f#1>>#1 to be supplied.\par}%
 \fi%
 \csname r@fnum#1\endcsname}
\def\ifuncit@d#1{\expandafter\ifx\csname r@fnum#1\endcsname\relax}%
\def\newr@f#1{\global\advance\r@fcount by1%
    \expandafter\xdef\csname r@fnum#1\endcsname{\number\r@fcount}}

\let\r@fis=\refis			
\def\refis#1#2#3\par{\ifuncit@d{#1}
   \newr@f{#1}%
   \w@rnwrite{Reference #1=\number\r@fcount\space is not cited up to now.}\fi%
  \expandafter\gdef\csname r@ftext\csname r@fnum#1\endcsname\endcsname%
  {\writer@f#1>>#2#3\par}}

\def\ignoreuncited{
   \def\refis##1##2##3\par{\ifuncit@d{##1}%
     \else\expandafter\gdef\csname r@ftext\csname
r@fnum##1\endcsname\endcsname%
     {\writer@f##1>>##2##3\par}\fi}}

\def\r@ferr{\endreferences\errmessage{I was expecting to see
\noexpand\endreferences before now;  I have inserted it here.}}
\let\r@ferences=\references
\def\references{\r@ferences\def\endmode{\r@ferr\par\endgroup}}

\let\endr@ferences=\endreferences
\def\endreferences{\r@fcurr=0
  {\loop\ifnum\r@fcurr<\r@fcount
    \advance\r@fcurr by 1\relax\expandafter\r@fis\expandafter{\number\r@fcurr}%
    \csname r@ftext\number\r@fcurr\endcsname%
  \repeat}\gdef\r@ferr{}\endr@ferences}


\let\r@fend=\endpaper\gdef\endpaper{\ifr@ffile
\immediate\write16{Cross References written on []\jobname.REF.}\fi\r@fend}

\catcode`@=12

\citeall\refto		
\citeall\ref		%
\citeall\Ref		%

\font\tif=cmr7 scaled \magstep4
\rightline{iassns-hep-92-20}
\vskip-8truept
\rightline{McGill/92-48}
\vskip-8truept \rightline{hep-th/9211016}
\title{\tif Conformally invariant off-shell string physics}
\author{\rm Robert C. Myers\footnote*{rcm@physics.mcgill.ca}}
\affil{\rm Physics Department, McGill University
Ernest Rutherford Building
Montr\'eal, Qu\'ebec, CANADA H3A 2T8}
\author{\rm Vipul Periwal\footnote{${}^\dagger$}{vipul@guinness.ias.edu/%
vipul@iassns.bitnet}}
\affil{\rm The Institute for Advanced Study
Princeton, New Jersey 08540-4920}
\abstract{Using recent advances in the understanding
of non-critical strings,
we construct a unique, conformally invariant continuation
to off-shell momenta of Polyakov amplitudes
in critical string theory. Three-point amplitudes are explicitly
calculated. These off-shell amplitudes possess some unusual, apparently
stringy, characteristics, which are unlikely to be reproduced in a string
field theory.
Thus our results may be an indication that some fundamentally new
formulation, other than string field theory, will be required to extend
our understanding of critical strings beyond the Polyakov path integral.}

\bigskip
\centerline{PACS:11.17.+y}
\endtopmatter
\baselineskip=14truept

Polyakov's derivation of the connection between conformal anomalies
and the critical dimensions of string theories\pref{pol0}
elucidated a multitude of features of string physics,
gleaned piece-meal in pioneering work.
Space-time scattering amplitudes of string excitations are calculated
as correlation functions of vertex operators in a functional integral
over the metric on the string world-sheet, and the space-time
string configurations\pref{bose}:
$$
\bigg\langle \prod_i \int\!\dd^2\!z_i \sqrt g \,V_i(z_i)\bigg\rangle
\equiv
\int {\dD g\ \dD X\over \hbox{vol.}(\hbox{Diff})
\hbox{vol.}(\hbox{Weyl})}\ \exp(-S[g,X])
\ \prod_i \int\! \dd^2\!z_i \sqrt g\, V_i(z_i)
.\eqno(deaf)$$
The measure is divided by the `volume' of the symmetries of the
classical action $S\equiv (8\pi)^{-1}
\int\!\dd^2\!z \sqrt g g^{ab}\part_a X^\mu\part_b
X_\mu,$ with $\mu=1,\dots,D$---namely, diffeomorphisms and local
Weyl rescalings on the world-sheet. Choosing conformal gauge, $g_{ab}
\equiv\ee^{2\phi}\hg_{ab}(m)$, and fixing diffeomorphisms \`a la
Faddeev-Popov, these functional integrals reduce to
$$
\int\!\dd m~{\dD \phi\over\hbox{vol.}(\hbox{Weyl})}
{\dD X\ \ \hbox{Det}_{\ssc\rm FP}' \over\hbox{vol.(c.k.v.)}}
\ \exp(-S[\hg,X])
\ \prod_i \int\! \dd^2\!z_i \sqrt {\hat g(m)} V_i(z_i)\ \ .
\eqno(ordre)$$
Here,
c.k.v.~stands for the conformal Killing vectors that must be taken into
account if the world-sheet is a sphere or a torus,
and $\dd m$ denotes the measure for
integrating over moduli labelling distinct conformal equivalence classes
of metrics on surfaces with one or more handles.
Eqs.~\(deaf) and \(ordre) are actually only equivalent if
Weyl rescaling survives as a symmetry of the quantum path integral.
This requires that $D=26$ in order to cancel
the anomalous dependences on the Weyl field, $\phi$, in the
measure factor,
$\dD X\,\hbox{Det}_{\ssc\rm FP}'/\hbox{vol.(c.k.v.)}$.\pref{pol0}
Also, one must impose various {\it space-time}
 mass-shell and polarization/gauge
conditions on the external string states to avoid any anomalous Weyl
dependences from normal-ordering the vertex operators. Combined these
restrictions ensure that $\phi$ completely decouples
from on-shell correlation functions in critical string theory.
Then the integration
over the Weyl factor cancels against the volume of the
group of Weyl rescalings in the denominator
(\ie $\int\dD \phi/\hbox{vol.}(\hbox{Weyl}) \equiv 1$).

Therefore mass-shell conditions can be obtained from requiring Weyl
invariance.  It follows, in the Polyakov approach, that the
calculation of amplitudes for {\it off-shell} string states
requires the ability to compute correlation
functions of vertex operators with an anomalous Weyl dependence,
in the normalized measure $\dD\phi/\vw .$  The
purpose of the present Letter is to show that recent advances in
non-critical string theories\pref{pol,david,mmdk}
have rendered such computations
practicable.

Off-shell amplitudes are of great physical interest for string theories,
as they are for field theories.  They are essential
for the derivation of effective actions, \eg the derivation
of effective potentials for particles such as the tachyon and the
dilaton, for the derivation of measures for integrating over moduli
of space-time instantons in string theory\pref{thankgross}, and for the
calculation of hadronic form-factors when one attempts to interpret
certain aspects of quantum chromodynamics in terms of effective string
theories.

\def\sft{s.f.t.}
In string theory despite intensive investigations in the past, off-shell
amplitudes have proven
to possess a remarkable intransigence.
Space does not permit an
extended discussion of previous work here\pref{number2}, but we provide
a summary to put our work in perspective:
\item{1.} The first attempts\pref{ven}\ gave integral formul\ae\ for
off-shell extensions of the Veneziano amplitude that obeyed various
physical criteria such as crossing symmetry, vector dominance of
form-factors, Regge behaviour and current conservation.
\item{2.} String field theories (\sft s)\pref{sft}
naturally provide off-shell extensions.
Such extensions are not dual, since \sft\ amplitudes are
sums of Feynman diagrams constructed from certain building blocks of
fixed geometry, and independence from the conformal frame of these
building blocks holds
only when all external legs are on-shell.  Thus, off-shell
\sft\ amplitudes are not `stringy', and often possess spurious
singularities.  S.f.t. is an economical
way of extending a first-quantized understanding of strings, and the
fact that such an extension does not exhibit stringy properties
off-shell is no reason for immediate derogation.
\item{3.} Bardak\c ci, and Bardak\c ci and Halpern,\pref{bard}\
considered an off-shell extension while investigating spontaneous
symmetry-breaking in dual models.  They introduced a fictitious dimension,
with momenta in this dimension restricted to  $\pm 1.$  This
enabled them to preserve conformal invariance while computing tachyon
amplitudes at zero space-time momentum.  It will be evident in the
following that this work comes closest to the approach based on the
Polyakov functional integral that we pursue here.
\item{4.} Since Polyakov's work\pref{pol0}, attempts have been made to
compute amplitudes on surfaces with boundaries, with off-shell external
states specified as matter configurations on the boundaries.  Some of
this work\pref{double}
uses known mathematical results for surfaces with a reflection
symmetry, treating surfaces with boundaries in terms of their `double'
surfaces, with the boundaries as the curves left invariant under the
reflection symmetry.  The imposition of
physical boundary data, which is naturally
independent of the parametrization of the surface, is not treated
in these works.  Another approach\pref{phil}\ attempted to
compute the functional integral directly for a cylinder, but
neglected the Weyl dependence in the
integration over reparametrizations of boundary data.
It might be supposed that the point-like
states considered in Ref.~\cite{phil}\ should not suffer from this
problem\pref{subtle}.

\def\dtx{\!\dd^2\!x}
\def\dtz{\!\dd^2\!z}
Returning to our initial line of development, one may ask:
Why is the computation of correlation functions of $\phi$ with the
measure $\dD\phi/\vw$ difficult?  The problem resides in the
non-linearity of the Riemannian metric that defines $\dD\phi.$
The full metric $g_{ab}$ is used to define
the norm on infinitesimal changes in the conformal factor
$$(\delta\phi,\delta\phi) = \int\dtx\sqrt{g}(\delta\phi)^2=
\int\dtx \sqrt{\hat g}\ee^{2\phi}(\delta\phi)^2,$$
which then
explicitly depends on $\phi.$  Treating the functional integral over
$\phi$ as a standard quantum field theory requires a translation
invariant measure $\dD_{\ssc 0}\phi,$
defined by the norm,
$(\delta\phi,\delta\phi)_{\ssc 0} = \int\dtx \sqrt{\hat g}
(\delta\phi)^2.$
As shown by
Mavromatos and Miramontes, and, independently, D'Hoker and
Kurzepa\pref{mmdk}, these two measures are related in a remarkably simple
way
$$
\dD\phi = \dD_{\ssc 0}\phi \exp\left(S_{\ssc L}
- {\mu\over \pi}\int \dtz\,\ee^{\al\phi}\right),
\eqno(mesure)$$
where $S_{\ssc L} \equiv \int \dtz/(6\pi)
\bigg[\part\phi\bar\part\phi + {1\over
4} \sqrt {\hat g} \hat R\phi \bigg]$ and
$\mu$ is the `cosmological constant'. The latter coefficient remains
undetermined by their computation, but
$\al$ is explicitly fixed (see below).
This relation was originally conjectured\pref{david}
in the study of two-dimensional gravity coupled
to conformal matter in conformal gauge.  It is important to note that
the derivation of eq.~\(mesure) is mathematically entirely
independent of the rest of the functional integrals involved.  It is
valid in non-critical string theory, and equally valid in the context of
critical string theory.

The only assumption in the present work will be in treating
the correlation functions using the methods of conformal
field theory.  For non-critical strings, this approach has been verified
by comparison with the results of matrix model techniques.
The stress tensor deduced from $S_{\ssc L}$ is
$T_{\ssc L} = {1\over 6} \left[(\part\phi)^2-\part^2\phi\right],$
and it is easily checked that the central charge $c_{\ssc L} = 0.$ Thus
the total central charge (for matter, ghosts and now, Liouville field)
remains zero.
The weight of an exponential operator $\ee^{\be\phi}$ is
${3\over 2}\be(\be+{1\over 3}).$
Off-shell vertex operators $V_i$ of weight
$(\Del,\Del)$ are dressed the same way as matter operators in
non-critical string theories to produce (1,1) operators
$\exp(\beta_{\ssc\Del}\phi)V_i$ with
$$
\be_{\ssc\Del} = {1\over 6}\left[\sqrt{25-24\Del}-1\right].
\eqno(dress)
$$
This is the unique solution for $\be_{\ssc\Del}$ such that
$\Del=1\Leftrightarrow \be_{\ssc\Del}=0,$ which insures that in the
on-shell limit, these off-shell amplitudes reduce precisely to
the usual on-shell amplitudes.
Rather puzzling is the non-analyticity in this prescription
at $\Del={25\over 24}$, since there is
no obvious physical reason to restrict $\Del\le {25\over 24}$.
While one expects cuts in loop amplitudes in field theories,
it seems difficult to interpret this non-analyticity as arising from
similar physics. A better understanding
is certainly required to extend the applicability of our approach, but for
the present,
we will restrict our attention to $\Del\le {25\over 24}$.

The presence of the cosmological constant in eq.~\(mesure)
is important for defining the integration
over $\phi.$ Insertions of cosmological
constant interaction `cancel' Liouville momentum carried by the off-shell
vertex operators, and the background charge term in $S_{\ssc L}.$
However, the treatment of the complete action
is rather subtle[\cite{gl,pol2}]. Here, treating the
cosmological constant term as a perturbatively defined interaction,
we determine $\al= \be_{\ssc \Del=0} = \ttt.$
One could consider the other branch of the square root,
which gives $\al=-1, $ but $\al=\ttt$ may be preferred since then
this interaction
can be interpreted as a zero-momentum tachyon, hence as obtained
from the off-shell continuation of a physical state.
Also if used as the area operator of the quantum
theory, a vanishing area results in the limit $\phi\rightarrow -\infty,$
in accord with classical expectations.

\def\tg{{3\gam+1\over2}}

Explicit computations can be performed on the two-sphere,
using the ideas of Goulian and Li\pref{gl}
to perform the integral over the constant zero-mode, $\phi_{\ssc 0}.$
The classic calculation of Dotsenko and Fateev\pref{dotf} can then be used
to compute the
resulting correlation function, with appropriate analytic continuations
along the way\pref{gl}.  The zero-mode integral is
$$\int d\phi_{\ssc 0} \exp\left({1\over 6}\phi_{\ssc 0} -{C}
{\ee}^{\ttt\phi_{\ssc 0}}
\right) \exp(\gam\phi_{\ssc 0})
= {3\over 2}\Gamma\left({1\over 2}(3\gam+1)\right) {C}^{-{1\over
2}(3\gam+1)},$$
where $\gam \equiv \sum\beta_i\equiv \sum \be(\Del_i),$
and ${C} \equiv (\mu/\pi)\int \dd^2 z \exp({2\over 3}\tilde\phi),$
with $\int \dd^2 z \tilde\phi=0$.
The (not yet normalized) amplitude is now
$$\Big\langle \prod_i \int \dd^2z_i\,\ee^{\be_i\phi} V_i(z_i)\Big\rangle
= {3\over 2} \Gamma(-s)
\prod_i\int\dd^2z_i \Big\langle{C}^s\prod_j \ee^{\be_j
\tilde\phi}(z_j)\Big\rangle_{L}\Big\langle \prod_kV_k(z_k)\Big\rangle_m,$$
where $s\equiv -{1\over 2}(3\gam+1),$ and the subscript $L(m)$ stands
for Liouville (matter) expectation values.
For three-point functions and positive integer
values of $s$, these correlations were treated by
Dotsenko and Fateev\pref{dotf}. Choosing three tachyon operators,
$V_j=\exp(ik_j^\mu X_\mu)$, and
fixing their positions
$\{z_1,z_2,z_3\}$ at $\{0,\infty, 1\}$, yields
$$
{\cal A}=
\hbox{$3\over 2$} \mu^s\Gamma\left(-s\right)\Gamma\left(s+1\right)
\Del(\hbox{$1\over 3$})^{s}\prod_{i=0}^{3}
\prod_{k=0}^{s-1}\Del(1+2\beta_i+\hbox{$2\over 3$}k)\ .
$$
Here $\Del(z) \equiv \Gam(z)/\Gam(1-z),$ and we have defined
$\be_{\ssc 0}\equiv -{1\over 6},$ but $\gam=\sum_{i=1}^3\be_i.$
Using the ideas of Ref.~\cite{gl}, the above formula can be continued
to the following expressions:
$$\eqalign{{\cal A} =
&\left[\mu\Del\left(\hbox{$1\over 3$}\right)\right]^{-\tg}
\Gam\left(\hbox{$1+3\gam\over2$}\right)
\Gam\left(\hbox{$1-3\gam\over2$}\right)\cr & \times
\left(\hbox{$2\over 3$}\right)^{\gam-{2\over 3}}
\prod_{i=0}^3\prod^{\gam-\ttt}_{p=0} \Del(1-3\be_i + \hbox{$
3\over 2$}p)\cr
{\rm or}\qquad &\times
\left(\hbox{$2\over 3$}\right)^{5\gam+4}
\prod_{i=0}^3 \Del(2\be_i-\gam)
\prod^{\gam}_{p=0} \Del(1-3\be_i + \hbox{$3\over 2$}p)\cr
{\rm or}\qquad &\times
\left(\hbox{$2\over 3$}\right)^{9\gam+14}
\prod_{i=0}^3\Del(2\be_i-\gam)\Del(2\be_i-\gam-\hbox{$\ttt$})
\prod^{\gam+\ttt}_{p=0} \Del(1-3\be_i + \hbox{$
3\over 2$}p)
,\cr}\eqno(formu)$$
where $\gam$ must be such that
the upper limits of the products are integers. Combining all of these
formul\ae,
we have results which are valid for $\gam=n/3$ where $n$ is a positive
integer or zero. A more extensive description of the analytic continuations
above will appear elsewhere\pref{number2}.

It is useful to investigate the analytic structure of these amplitudes
when $\gam$ is held fixed.  Considering the ratio of two
such amplitudes (with the same values of $\gamma$), one finds
that the interesting dependence on $\be_i$ resides in, {\it e.g.},
$$
\prod_{i=1}^3 \Del(2\be_i-\gam)
\prod^{\gam}_{p=0} \Del(1-3\be_i + \hbox{$3\over 2$}p)\ .
$$
One finds poles and {\it zeroes}
depending on the value of $\be_i$ individually,
and $\gamma$ as well. Note that the restriction which arose in the
discussion of the dressings, $\Del\le {25\over 24}$, also constrains
$\beta_i\ge-{1\over6}$. For a fixed $\gamma$, this restricts the
number of poles and zeroes which actually occur.
A case of interest because the
particles can all go on-shell is $\gam=0$, where we find $\prod_{i=1}^3
\Del(1-3\be_i)\Del(2\be_i).$  This expression has poles where
$\beta_i\rightarrow1/3$ (\ie $k^2_i\rightarrow{4\over3}$),
and no zeroes---in particular, it remains finite as
$\be_i\rightarrow 0.$
A striking feature of the
amplitudes is the presence of poles that are not accounted for by
excitations in the matter sector (even if combined with the ghost sector).
They may indicate the presence of
excitations that are entirely stringy in nature.

Independent of the existence of new poles, the fact that the amplitudes
have products which have upper limits determined by $\gam$ is something
entirely unlike the amplitudes one obtains from a field theory.  In
field theories, the off-shell character of the amplitude is a function
of individual external states.  Here, one can obtain the value $\gam=0$
when all external states are on-shell, {\it or} if they are off-shell.
It is difficult to imagine how this $\gam$ dependence could be reproduced
in a string field theory. Thus our results may be an indication
that some fundamentally new framework, other than string field theory,
will be required to extend our understanding of
critical string theory beyond the
Polyakov path integral. Of course, even though our present knowledge
of string theory is derived almost entirely from
the Polyakov functional integral, it is not possible to exclude the
possibility that the Polyakov approach is just a recipe for on-shell
calculations.

An important feature which distinguishes our amplitudes from
those of non-critical string is  the factor $\vw$ in the denominator
of eq.~\(ordre). The computation of the Weyl volume is subtle.
Ref.~\cite{gl} gives a prescription which uses eq.~\(formu) with $\gam=2$
and $\beta_i=2/3$ to give a result for the two-sphere
(which actually vanishes).
At tree-level it is possible to
evade a direct computation of the Weyl volume by considering ratios of
amplitudes.  On higher genus surfaces, the presence of this factor
ensures that the Weyl field does not show up in
any counting of states via degenerations.
In particular, the dependence on the moduli in $\dD \phi$
is precisely cancelled by the
denominator, unless there are off-shell vertex operators present.
Note then that in eq.~\(ordre), $\dd m$ and $\dD \phi/\vw$ must be
explicitly ordered as given.

In conclusion, we have shown in this Letter that the effort expended on
the study of non-critical strings in somewhat unphysical contexts has
important physical consequences in critical string theories.
It follows as well that any new future insights into
non-critical string physics, or into quantum Liouville theory, will translate
directly into further insights into off-shell critical string physics.
 There are
a great many physical questions that become accessible in our approach
to off-shell string physics.
Above we have only considered simple exponential dressings, but one can
also find many new (1,1) primary fields with oscillator contributions
(\eg $\part\phi$) which will couple in amplitudes. Some of these may
account for longitudinal polarizations which only couple off-shell.
It is possible to derive explicit formul\ae\ for
four-point functions with mild kinematic restrictions, and of
course, a supersymmetric extension of these ideas is immediate.
Extending eq.~\(formu) to arbitrary values of $\gam$ is required,
as is an understanding of the non-analyticity in eq.~\(dress).
The calculation of the effective action requires care.
Computations for zero-momentum tachyons do not contain
expected dilaton exchange singularities, and yield a {\it non-analytic}
tree-level effective potential,
$$\Gamma(T) \sim 3\,T^{1\over 3}-T.$$
An extended treatment of these issues is
in preparation\pref{number2}.

Acknowledgements: We thank J.~Distler, Vl.~Dotsenko, J.~Minahan, P.~Nelson,
and especially D. Gross, for helpful conversations.
R.C.M. was supported by NSERC of Canada, and Fonds FCAR du Qu\'ebec.
V.P. was supported by D.O.E. grant DE-FG02-90ER40542.
\vfill\eject

\references

\refis{pol0} A.M. Polyakov, {\sl Phys. Lett.} {\bf 103B} (1981) 207,
211

\refis{number2} R.C. Myers and V. Periwal, IAS/McGill preprint
iassns-hep-92-36, to appear

\refis{bose} We restrict ourselves to the bosonic string in this Letter.

\refis{subtle} Normalizing the boundary diffeomorphism volume would
appear to create problems of interpretation for factorized amplitudes.
See also: F. Fucito, {\sl Phys. Lett.} {\bf 193B} (1987) 233

\refis{thankgross} We are grateful to D. Gross for pointing this out.

\refis{mmdk} N.E. Mavromatos and J.L. Miramontes, {\sl Mod. Phys. Lett.}
{\bf A4} (1989) 1847; E. D'Hoker and P.S. Kurzepa, {\sl Mod. Phys. Lett.}
{\bf A5} (1990) 1411; E. D'Hoker, {\sl Mod. Phys. Lett.} {\bf A6} (1991) 745

\refis{gl} M.~Goulian and M.~Li, {\sl Phys. Rev. Lett.} {\bf 66} (1991)
2051; see also A. Gupta, S.P. Trivedi and M.B. Wise, \nuc 340,1990,475

\refis{dotf} Vl.S.~Dotsenko and V.~Fateev, {\sl Nucl. Phys.} {\bf B240}
(1984) 312, {\bf B251} (1985) 691

\refis{pol} A.M. Polyakov, {\sl Mod. Phys. Lett.} {\bf A2} (1987) 893;
V.G. Knizhnik, A.M. Polyakov and A.B. Zamolodchikov, {\sl
Mod. Phys. Lett.} {\bf A3} (1988) 819

\refis{pol2} A.M. Polyakov, {\sl Mod. Phys. Lett.} {\bf A6} (1991) 635

\refis{david} F. David,{ \sl Mod. Phys. Lett.} {\bf A3} (1988) 1651;
J. Distler and H. Kawai, {\sl Nucl. Phys.} {\bf B321} (1989) 509

\refis{ven} Some references:
M. Bander, {\sl Nucl. Phys.} {\bf B13} (1969) 587;
A. Ademollo and E. Del Giudice, {\sl Nuovo Cim.} {\bf 63A} (1969) 639;
R.C. Brower and M.B. Halpern, {\sl Phys. Rev.} {\bf 182} (1969) 1779;
R.C. Brower and J.H. Weis, {\sl Phys. Rev.} {\bf 188} (1969) 2486, 2496,
{\bf D3} (1971) 451;
R.C. Brower, A. Rubl and J.H. Weis, {\sl Nuovo Cim.} {\bf A65} (1970) 659;
C. Rebbi, {\sl Lett. Nuovo Cim.} {\bf 1} (1971) 967;
E. Napolitano and S. Sciuto, {\sl Lett. Nuovo Cim.} {\bf 1} (1971) 1095;
C. Bouchiat, J.-L. Gervais and N. Sourlas, {\sl Lett. Nuovo Cim.} {\bf
3} (1972) 767;
B. Hasslacher and D.K. Sinclair, {\sl Lett. Nuovo Cim.} {\bf 4} (1972) 515;
M. Ademollo and J. Gomis, {\sl Nuovo Cim.} {\bf 4} (1971) 299;
P.V. Landshoff and J.C. Polkinghorne, \nuc 19,1970,767;
J.H. Schwarz, {\sl Nucl. Phys.} {\bf B65} (1973) 131;
J.H. Schwarz and C.C. Wu, {\sl Nucl. Phys.} {\bf B72} (1974) 337;
E. Corrigan and D.B. Fairlie, {\sl Nucl. Phys.} {\bf B91} (1975) 527;
K. Bardak\c ci and M.B. Halpern, {\sl Nucl. Phys.} {\bf B73} (1974) 295;
N.V. Borisov, M.V. Ioffe, N.A. Liskova and M.I. \'Eides, {\sl Sov. J.
Nucl. Phys.} {\bf 33} (1981) 876; M.B. Green and J. Shapiro,
{\sl Phys. Lett.} {\bf 64B} (1976) 454; M.B. Green, \nuc 116,1976,449,
{\sl Phys. Lett.} {\bf 65B} (1976) 432, {\bf 69B} (1977) 89,
\nuc 124,1977,454, {\sl Phys. Lett.} {\bf 201B} (1988) 42,
{\bf 266B} (1991) 325, {\bf 282B} (1992) 380

\refis{sft} Some references: 
J.H. Sloan, \nuc 302,1988,349 ; S. Samuel, \nuc 308,1988,{285, 317};
O. Lechtenfeld and S. Samuel, \nuc 308,1988,{361, {\bf B310} (1988) 254};
V.A. Kostelecky and S. Samuel, {\sl Phys. Lett.} {\bf 207B} (1988) 169;
R. Bluhm and S. Samuel, \nuc 323,1989,{337, {\bf B325} (1989) 275};
B. Sathiapalan, {\sl Phys. Lett.} {\bf 206B} (1988) 211;
D.Z. Freedman, S.B. Giddings, J. Shapiro and C.B. Thorn, \nuc
298,1988,253;
A. Ukegawa, {\sl Phys. Lett.} {\bf 261B} (1991) 391;
J. Feng, \nuc 338,1990,459; K. Sakai, {\sl Prog.  Theor.
Phys.} {\bf 80} (1988) 294

\refis{phil} A. Cohen, G. Moore, P. Nelson and J. Polchinski, {\sl Nucl.
Phys.} {\bf B267} (1986) 143, {\bf B281} (1987) 127, {\sl Phys. Lett.}
{\bf 169B} (1986) 47;
W.I. Weisberger, {\sl Nucl. Phys.} {\bf B294} (1987) 113;
A.N. Redlich, {\sl Phys. Lett.} {\bf 205B} (1988) 295;
C. Varughese, SUNY (Stony Brook) thesis (1989), UMI-90-11440-mc;
Z. Jaskolski, {\sl Comm. Math. Phys.} {\bf 139} (1991) 353;
M.A. Martin-Delgado and J. Ramirez Mittelbrunn, {\sl Int. J. Mod. Phys.}
{\bf A6} (1991) 1719

\refis{double} M. Bershadsky, {\sl Sov. J. Nucl. Phys.} {\bf 45} (1987) 925;
S.K. Blau, M. Clements, S. Della Pietra, S. Carlip and V. Della Pietra,
\nuc 301,1988,285;
J. Bolte and F. Steiner, {\sl Nucl. Phys.} {\bf B361} (1991) 451

\refis{bard}
K. Bardak\c ci,
{\sl Nucl. Phys.} {\bf B68} (1974) 331, {\bf B70} (1974) 397;
K. Bardak\c ci and M.B. Halpern,
{\sl Phys. Rev.} {\bf D10} (1974) 4230,
{\sl Nucl. Phys.} {\bf B96} (1975) 285;
K. Bardak\c ci, {\sl Nucl. Phys.} {\bf B133} (1978) 297

\endreferences
\end

This is probably useful since we can consider $\ga=0$
to examine amplitudes in the vicinity of being on-shell.
One can also use $\ga=2$ to define the Weyl volume from
a three-point amplitude of zero-momentum tachyons. Using the rearrangement
formula, the un-normalized amplitude for three tachyons is
$$\eqalign{
 A={4\over9}{\pi\over\sin{3\pi\over2}(\ga+1)}&(\ga+1)\mu^{-{3\ga+1\over2}}
\Del({1\over3})^{-{3\over2}(\ga+1)}\prod_{l=1}^3\Del(1+2\be_l)\cr
&\prod_{p=0}^\ga\Del({1\over2}+{3\over2}p)\prod_{l=1}^3
\Del({3\over2}p-3\be_l)\cr}$$

For the Weyl volume take $\be_1=\ttt=\be_2=\be_3$,
$$
 A_o={4\over9}{\pi\over\sin{9\pi\over2}}3\mu^{-{7\over2}}
\Del({1\over3})^{-{9\over2}}\Del(7/3)^3
\ \prod_{p=0}^2\Del({1\over2}+{3\over2}p)\Del({3\over2}p-2)^3 $$
Taking a limit $\be_l\rightarrow2/3$, regularizes the
product of the last terms:
$\Del(-2)\Del(-1/2)\Del(1)=-1$, but one still has from the
second-to-last terms: $\Del(1/2)\Del(2)\Del(7/2)=0$. Anyway defining
the Weyl volume $Z_\phi$ by $A_o=\pi^3\part_\mu Z_\phi$, one
finds
$$Z_\phi=-{8\mu^3\over15\pi^3}A_o\ \ .$$

Now for $\ga=0$, the un-normalized amplitude becomes
$$A=-(finite\ positive\ \#)\mu^{-1/2}\prod_{l=1}^3{\Gam(1+2\be_l)\over
\Gam(1+3\be_l)}{\Gam(-3\be_l)\over\Gam(-2\be_l)}\ .$$
Note this has a finite limit as $\be_l\rightarrow0$. The fact that this
result is finite while $A_o$ is zero would be consistent with our
expectations that an extra SL(2,C) volume arises on-shell. Note that
we are getting this volume/infinity not just on-shell but for $\ga=0$
-- more later. In this amplitude, we find
poles at $\be_l=(3n\pm1)/3$ and zeros at $\be_l=(2m+1)/2$
(but recall that we should restrict our attention to $\be_l\ge-1/6$)
-- so what?

More on SL(2,C): In general, if we ignore the $\be$ dependent stuff, there
is a ``volume'' factor that goes like $\Gam(3/2(\ga+1))\Gam(-(3\ga+1)/2)
\prod_{p=0}^\ga \Del((3p+1)/2)$. This term is finite for $\ga=0,1$, but
vanishes for $\ga\ge2$. As $\ga$ increases, it is actually vanishing more
strongly in the sense that you get more factors of $1/\Gam(-n)$.
Altogether, therefore
the picture of the SL(2,C) volume seems to be somewhat naive??

$\bullet$ There {\it may} exist a sensible semi-classical limit for the three
point amplitude at large $\ga$, but one has to figure out a way to
take $\ga,\be_l\rightarrow\infty$ which avoids running into any poles.
This may be required to justify/sanctify whatever analytic continuation
our results represent.

$\bullet$ One can also do a four point function for four tachyons in the
special cases that the fourth momentum is either zero or on-shell.
The latter case gives $A_4=A_3 X$ where $A_3$ is just the three-point
amplitude calculated above (for whatever $\ga$), and $X$ is the
Shapiro-Virasoro amplitude
$$ X=\pi{\Gam(1+k_1\cdot k_4)\over\Gam(-k_1\cdot k_4)}
{\Gam(1+k_2\cdot k_4)\over\Gam(-k_2\cdot k_4)}
{\Gam(1+k_3\cdot k_4)\over\Gam(-k_3\cdot k_4)}$$
The poles here are in $k_l\cdot k_4$ being negative integers.
These can be expressed in terms of the usual kinematic variables
like $t=(k_1+k_4)^2$ as poles at $t+3(\be_1+{1\over6})^2=
\{2,0,-2,\ldots\}+{1\over12}$. So poles do shift or get extended
off-shell (contrary to any arguements I may have made)!

$\bullet$ The last interesting thing that I did was consider a massless
tensor vertex operator of the form:
$$V_T=\Lam_{\mu\nu}(\part X^\mu + {ik^\mu\over1+3\la}\part\phi)
(\bar\part X^\nu+ {ik^\nu\over1+3\la}\bar\part\phi)exp(ik\cdot X+\la\phi)$$
where $\la=(\sqrt{1-12k^2}-1)/6$. So you have off-shell $\phi$ dressing but
also a non-transverse dressing as well. In a 1 tensor -- 2 tachyon amplitude,
the answer is $A\,Y(\Lam,k)$ where $A$ is again the same three point
amplitude as above, and $Y(\Lam,k)=-\Lam_{\mu\nu}P^\mu{}_\rho P^\nu{}_\ga
k_2^\rho k_2^\ga$ is the kinematic factor. $P^\mu{}_\rho=\delta^\mu{}_\rho-
k_3^\mu k_{3\rho}/(k_3{}^2)$ is a transverse projection operator for the
tensor state with momentum $k_3$ (\ie $k_3\cdot P=0=P\cdot k_3$). So we
get the same kinematic factor as the standard on-shell amplitude, up to
the projection operators which come out of the off-shell calculation.
Note that the projection operators are singular on-shell (\ie $k_3{}^2=0$),
and I'm not sure how to evaluate the amplitude at that point.

\end

\refis{siegel} W. Siegel, {\sl Phys. Lett.} {\bf 151B} (1985) 391, 396

\refis{seib} N. Seiberg, {\sl Prog. Theor. Phys. Suppl.}
{\bf 102} (1990) 319